\documentclass[aps,twocolumn,showpacs]{revtex4}
\usepackage{graphicx}
\usepackage{bm}
\usepackage{amsmath}

\begin{document}

\title{Theory for supersolid $^4$He }
\author{Xi Dai$^1$, Michael Ma$^{2}$, and Fu-Chun Zhang$^{1,2,3}$}
\date{\today}

\affiliation{$^1$Department of Physics, University of Hong Kong, Hong Kong\\
$^2$Department of Physics, University of Cincinnati, Cincinnati, Ohio 45221,USA\\
$^3$Department of Physics, Zhejiang University, Hangzhou, China}
\date{\today}

\begin{abstract}
Although both vacancies and interstitial have relatively high
activation energies in the normal solid, we propose that a lower
energy bound state of a vacancy and an interstitial may facilitate
vacancy condensation to give supersolidity in $^{4}$He . We use a
phenomenological two-band boson lattice model to demonstrate this
new mechanism and discuss the possible relevance to the recently
observed superfluid-like, non-classical rotational inertial
experiments of Kim and Chan in solid $^{4}$He. Some of our results
may also be applicable to trapped bosons in optical lattices.
\end{abstract}

\maketitle

\narrowtext Recently Kim and Chan have reported observation of
superfluid-like, non-classical rotational inertial (NCRI) behavior
in solid $ ^{4}$He, both when embedded in Vycor glass~\cite{kim1}
and in bulk $^{4}$He~ \cite{kim2}. Their experiments have revived
great interest in supersolidity with both crystalline and
superfluid orderings in helium. The possibility of a supersolid
phase in $^{4}$He was theoretically proposed by Chester~\cite
{chester}, Leggett~\cite{leggett}, Saslow~\cite{saslow} and by
Adreeve and Lifshitz~\cite{andreev} in 1970's. Adreeve and
Lifshitz proposed Bose condensation of vacancies as the mechanism
for supersolidity. Chester speculated that supersolidity cannot
exist without vacancies and/or interstitials~\cite{chester}, a
claim made more rigorous recently by Prokofev and Svistunov
\cite{russian}. Experiments and more sophisticated microscopic
calculations have, however, provided constraints to any theory for
supersolid in $^{4}$He. The NMR experiments \cite{guyer,meisel} on
solid $^{3}$He rule out a non-negligible zero point vacancy
concentration. The energy of a vacancy in solid $^{4}$He was
estimated to be about $10K$ by the x-ray scattering
experiment~\cite {x-ray} and to be $15K$ in a theoretical
calculation \cite{chaudhuri}. The energy of a pure interstitial
state has recently been calculated to be about $48\pm
5K$~\cite{ceperley}. Assuming that the observed supersolidity is a
genuine bulk phenomenon, the quandary is thus how defects with
such high activation energies can condense at low temperature.

In this Letter, we propose a possible solution to this quandary.
In addition to vacancies and interstitials, there is a third type
of defects, with relatively low excitation energy, which
corresponds to a bound state of a vacancy and an interstitial,
henceforth called an ''exciton''. While such excitons do not carry
mass and will not contribute to supersolid phenomena like those
observed by Kim and Chan~\cite{kim1,kim2}, they can facilitate
supersolidity by two mechanisms. First, vacancies can condense
above a background of excitons easier than above the defect free
(DF) normal state, so that the condensation energy can more than
compensate for the exciton excitation energy. Second, in a
background formed of a coherent mixture of the DF state and the
exciton, the effective kinetic energy of vacancies and
interstitials can be enhanced due to constructive interference
between hopping processes involving the DF state and those
involving the exciton. The essence of our theory is that while,
consistent with all known experiments, the normal solid state is
the DF state, the supersolid state results from condensation of
vacancies and/or interstitials about a defect rich background of
excitons. This physics is shown quantitatively using a
phenomenological two-band lattice boson model to represent the
defects in solid $^{4}$He. Using mean field theory (MFT), we show
that superfluidity in solid helium can exist in parameter regimes
qualitatively consistent with all the known experiments and
microscopic calculations. We will argue that the key results will
hold beyond MFT, and indeed are rendered more robust by inclusion
of quantum fluctuations. Because of the ''vacuum switching''
between the normal and supersolid states, the transition at zero
temperature is generically first order.

We start with the lattice as defined by the periodicity of crystalline
helium. On each lattice site, we consider two single-particle localized
states. The lower energy state ($a$-state), with energy $-\epsilon _{a},$
has its maximum on the lattice site. The other state ($b$-state), with a
higher energy $-\epsilon _{a}+\Delta $, is less localized and has maxima
distributed with hcp symmetry away from the lattice site. Because of the
strongly repulsive cores, each of this state can hold at most one $^{4}$He
atom, i.e. for each state, helium behaves as hard core bosons. On the other
hand, because of the spatial separation between the $a$ and $b$ states, an
atom in the $a$-state repels one in the $b$-state with a large but weaker
strength $U.$ An atom on one site can tunnel to a neighboring site from $a$
to $a$-state, $b$ to $b$ -state, and $a$ to $b$ state with hopping
amplitudes $t_{a}$, $t_{b}$, and $t_{ab}$ respectively, taken all to be real
and non-negative. In the DF state, we have one helium atom occupying the $a$%
-state on each lattice site. Due to the hard core condition, the $^{4}$He
atoms in this state are immobile, and this state is a normal solid, which is
consistent with the assumption that defects are necessary for supersolid. To
study defects, it is convenient to consider the DF state as the defect
vacuum state, from which vacancies and interstitials can be
created. The term vacuum will always refer to the DF state henceforth in
this article. Let $a_{i}^{\dag }$ be the creation operator for the vacancy
by removing a helium atom from this vacuum state at the site $i$, and $%
b_{i}^{\dag }$ the creation operator for the interstitial by creating an
atom to the $b$-state on the site. Note that the interstitial defined this
way can be viewed as a quantum generalization of the classical interstitial.
The hard core conditions specified above implies these operators are also
hard core boson operators. Our model Hamiltonian for defects in solid helium
is then,

\begin{eqnarray}
H &=&\sum_{j}\epsilon _{a}n_{j,a}+\epsilon _{b}n_{j,b}-Un_{j,a}n_{j,b}
\nonumber \\
&-&\sum_{<ij>}(t_{a}a_{i}^{\dag }a_{j}+t_{b}b_{i}^{\dag
}b_{j}+t_{ab}a_{i}^{\dag }b_{j}^{\dag }+h.c.)  \label{original_H}
\end{eqnarray}
In the above equation, $n_{j,a}=a_{j}^{\dag }a_{j}$, $n_{j,b}=b_{j}^{\dag
}b_{j}$, and $\epsilon _{a}$ (previously defined) and $\epsilon _{b}$ are
the energy cost to create a helium vacancy and an interstitial respectively.
The repulsion $U$ between He atoms translates into an attraction of strength
$U$ between a vacancy and an interstitial on the same site. The energy of a
localized exciton is $\Delta =\epsilon _{a}+\epsilon _{b}-U$. For
simplicity, in what follows we will consider $t_{ab}=0$.

To illuminate the effects of excitons, let's first discuss the
non-interacting case $U=0$. In this limit, the vacancies and interstitials
are decoupled. For $\epsilon _{b}\rightarrow \infty $, our model is reduced
to the vacancy model proposed by Andreeve and Lifshitz \cite{andreev}.
Let $z(=12)$ be the number of the nearest neighbors in a hcp lattice, the
onset of Bose condensation of vacancies (interstitials) is given exactly by $%
\epsilon _{a(b)}-zt_{a(b)}=0^{-}$, which coincides with having
zero activation energy for a single defect, a condition that is
not supported by experiments\cite{x-ray} and theoretical
estimates\cite{chaudhuri,ceperley} for $^{4}$He. This is the
quandary we stated in the introduction.

We now consider the case $U>0$. In this case, a vacancy and an interstitial
tend to bind together to form a local exciton. If the exciton energy $\Delta
$ is small (large $U)$, the presence of the excitons in the ground state
enhances the kinetic energy of the vacancy (or interstitial), which may lead
to the condensation of these defects. Since we are primarily interested in
the large $U$ case, it is important to treat the on-site attractive
interaction accurately. In order to do so, we use the single-site mean field
approximation (MFA) by decoupling the kinetic terms as
\begin{eqnarray}
a_{i}^{\dag }a_{j}&\approx &\bar{a}(a_{j}+a_{i}^{\dag })-\bar{a}^{2},
\nonumber \\
b_{i}^{\dag }b_{j}&\approx &\bar{b}(b_{j}+b_{i}^{\dag })-\bar{b}^{2}
\end{eqnarray}
where the spatially uniform Bose condensation order parameters $\bar{a}
=<a_{i}>$ and $\bar{b}=<b_{i}>$ are determined self-consistently. The single
site mean field Hamiltonian for the hard core bosons are then solved
exactly. This MFA gives the correct exact conditions for onset of
superfluidity at $U=0$.

\begin{figure}[tbp]
\begin{center}
\includegraphics[width=9.2cm,angle=0,clip=]{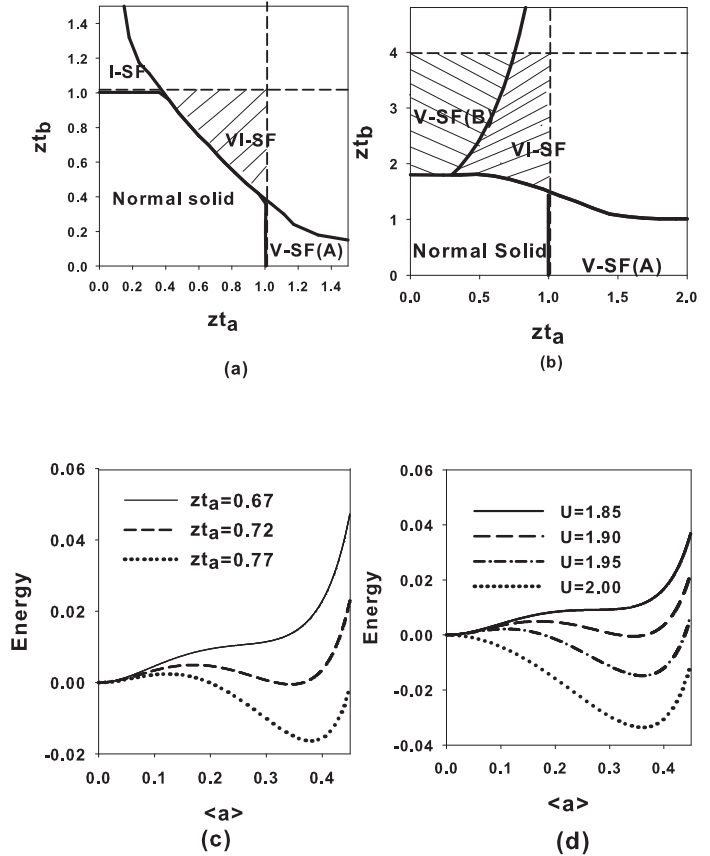}
\end{center}
\caption{(a) Mean field theory phase diagram of Hamiltonian (1) in the
symmetric case with $\epsilon_a=\epsilon_b=1$ and $U=1.9$. V-SF (I-SF):
vacancy (interstitial) superfluid phase. (b) Same as in (a) in the
asymmetric case $\epsilon_a=1$, $\epsilon_b=4$ and $U=4.8$. See Fig. 2 for
the snapshots of these phases in real space. (c)The variational energy of
the symmetric model with $\epsilon_a=\epsilon_b=1$ and $U=1.9$ as the
function of the SF order parameter for $t=t_a=t_b=0.056$ (normal solid
phase), $t=0.06$ (at the phase boundary) and $t=0.064$ (VI-SF phase). (d)
Same as in (c) with $t_a=t_b= 0.06$ for various values of $U$. }
\label{fig1}
\end{figure}

\begin{figure}[tbp]
\begin{center}
\includegraphics[width=8cm,angle=0,clip=]{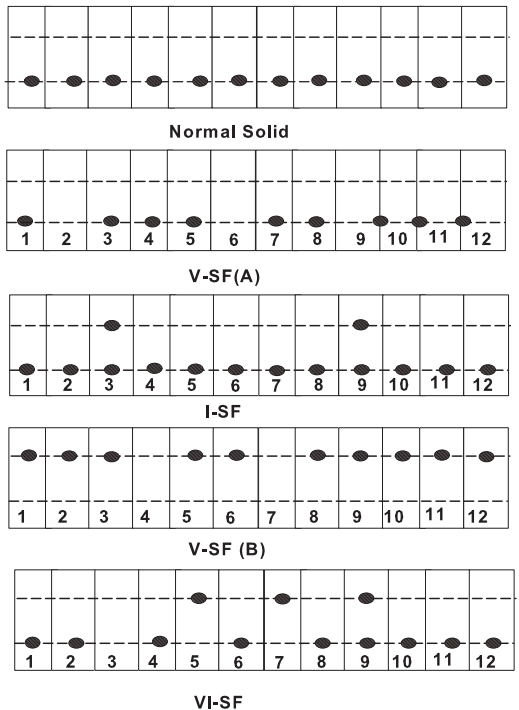}
\end{center}
\caption{Schematic illustration of snapshots of five phases shown in fig.\ref
{fig1}(a) and (b) in terms of the occupation of helium atoms at the lattice
sites. A lattice site is represented by a numbered block. A solid dot
represents a helium atom. At each site, emptiness represents a vacancy,
occupation at the lower (higher) level represents a vacuum (exciton), and
occupation at the both levels represents an interstitial. }
\label{snapshot}
\end{figure}

The ground state phase diagrams obtained within the MFA in the parameter
space $t_{a}$ and $t_{b}$ are shown in Fig. 1(a) for $\epsilon _{a}=\epsilon
_{b}=1$, $U=1.9$ and in Fig. 1(b) for $\epsilon _{a}=1$, $\epsilon _{b}=4$,
and $U=4.8$. They represent a more symmetric and a strongly asymmetric cases
respectively. Within the MFA, we found five different phases, characterized
by the order parameters $\bar{a}$ and $\bar{b}$ together with the vacuum,
vacancy, interstitial, and exciton defect densities $n_{0},n_{V}$, $n_{I},$
and $n_{ex}$. They are (1) The normal DF solid phase. (2) The vacancy
superfluid phase (V-SF(A)), where only the vacancies condense ($\bar{a}\neq
0 $) above the DF background ($n_{I}=n_{ex}=0).$ This phase is the same as
the vacancy state of Andreeve and Lifshitz. (3) The corresponding
interstitial superfluid phase (I-SF). (4) An alternative vacancy superfluid
phase (V-SF(B)), where the vacancies condense above a background of
excitons. (5) The VI-SF phase, where we have both vacancies and
interstitials condensing above a background of a coherent mixture of the
vacuum and the exciton. A snapshot of each of these phases in terms of the
occupation of helium atoms in real space is illustrated in Fig. \ref
{snapshot}.

To appreciate the phase diagrams and the new physics arising from
the interaction $U$, we show the phase boundaries of the
non-interacting model ($ U=0$) by dashed lines in Fig. 1(a) and
(b). As remarked earlier, these boundaries coincide with the
vanishing of the activation energy of an isolated vacancy or
interstitial. Since $U$ plays no role if only vacancies or only
intersititials are present, these are also the phase boundaries
for normal solid to V-SF(A) or I-SF transition even for $U>0$.
Thus, for such SF states, their corresponding normal (i.e.
uncondensed) states are not the DF state but contain a finite
density of defects, which is not the case for supersolid $^{4}$He.
The most striking of our results is that the interplay between
vacancies (and/or intersitials) and exciton defects can lead to SF
even in parameter regimes where the normal state is stable against
the generation of uncondensed defects (hatched areas in the
figures). We propose this as the reconciliation between activated
defect behavior at high temperature ($T$) and supersolidity at low
$T$ in $^{4}$He. In the hatched regime, the DF state is metastable
but not the global minimum energy state. In Fig. 1(c) and 1(d), we
illustrate this by plotting the variational energy $E(\bar{a})$
based on the MFA ~\cite{variation} for the symmetric model with
$t_{a}=t_{b}=t$, so that $\bar{a}=\bar{b},$ with increasing $t$
(Fig.1(c)) and $U$ (Fig.1(d)). The existence of two minima are
clearly seen. With increasing $t$ or $U$, the Bose condensed state
becomes lower in energy than the normal state in the hatched
areas. The transition is first order due to ''vacuum switching'':
the normal state at $\bar{a}=0$ is the DF state with no excitons
($n_{ex}=0$)$,$ while the background that vacancies and
intersitials condense above to form the VI-SF is the one with a
finite density of excitons ($n_{ex}\neq 0).$ In what follows we
discuss in more details the physics of the asymmetric and
symmetric cases, focusing on the parameter regimes of the hatched
areas.

Due to the high He atom density and strong repulsive cores, solid
$^{4}$He should fit the very asymmetric case with $\epsilon
_{b}/\epsilon _{a}>>1$ in our model (Fig. 1(b)). Because the
$b$-state is less localized compared to the $a$-state and because
of the exponential dependence of the overlap integral on state
size, we expect $t_{b}>>t_{a}.$ For small $t_{a},$ the phase in
the hatched region is the V-SF(B) phase. To understand how this
phase comes about, we consider first $t_{a}=$ $0.$ Here, the
activation energies of a vacancy and an exciton are $\epsilon
_{a}$ and $\Delta $ respectively independent of $t_{b},$ so the DF
state is locally stable. Since $t_{a}=0,$ vacancies cannot hop,
and there is no possibility of vacancy condensation above the DF
state. However, if we take a background of excitons on every site,
then vacancies can now hop with amplitude $t_{b},$ and the exciton
state can be unstable with respect to a vacancy condensation that
gives $\bar{b} \neq 0.$ The onset of this V-SF(B) instability is $
zt_{b}=\epsilon _{a}- \Delta .$ For $t_{b}$ greater than this value, $E(\bar{%
b})$ has a double minima behavior similar to that shown in Fig
1(c) for the symmetric case, with $E(\bar{b}=0)$ initially the
lower energy. However, as $ t_{b}$ increases (but still less than
the value necessary for spontaneous generation of interstitials),
the vacancy condensation energy can become large enough to
overcome the required vacancy and exciton activation energies
given by $n_{V}\epsilon _{a}+(1-n_{V})\Delta ,$ and a first order
transition from the normal DF state to the V-SF(B) state occurs.
Note that the transition is accompanied by a ''vacuum switching'',
in that the vacancies are condensing not above the DF state, but
above the exciton state. The V-SF(B) has $\bar{b}\neq 0,$ and
$n_{V}+n_{ex}=$ $1.$ As $t_{a}$ increases, it becomes advantageous
to mix in some vacuum component to allow vacancy hopping through
$t_{a}$, so that both $\bar{a}$ and $\bar{b}$ are non-zero.
Depending on the value $\epsilon _{b},$ some interstitial
condensation will also occur. The V-SF(B) phase then makes a
second order transition into the VI-SF, characterized by both
$\bar{a}$ and $\bar{b}\neq 0,$ and $n_{V,}$ $n_{I,}$ $n_{ex}$ all
$\neq 0.$ We will discuss this phase further in the symmetric
case, where it will feature more prominently. In this strongly
asymmetric case, whether we take the SF state to be the V-SF(B) or
the VI-SF state, we have $n_{V}>>n_{I}.$ Thus, a prediction of our
theory is that while the normal state is a commensurate solid, the
supersolid will have incommensurate density, which can be
confirmed by neutron scattering experiments in the supersolid
phase. Furthermore, because of the presence of excitons in the
supersolid (i.e. the He atom resides in the less localized $b-$
state rather than the $a-$ state, or some linear combination of
the two), the local He density in a unit cell will change with the
transition into the supersolid. This can be confirmed from the
form factor of neutron scattering or with a local probe.

Next we look at the symmetric case. Although this probably does
not describe solid $^{4}$He, it may be applicable to a system of
trapped bosons in an optical lattice, where the periodic potential
is imposed externally rather than internally generated. In such a
system, the ratio $\epsilon _{b}/\epsilon _{a}$ can be tuned by
tuning the optical potential. The interesting phase here is the
VI-SF phase in the hatched region. Again, the key question is what
causes the additional non-trivial SF solution since neither
vacancy nor the interstitial alone can condense at $T=0$. Unlike
the asymmetric case, where $n_{I}$ in this phase basically plays
no role, and indeed $\rightarrow 0$ as $\epsilon _{b}\rightarrow
\infty ,$ here, $ n_{0},n_{V},n_{I},n_{ex}$ are all non-negligible
in this SF phase. Within the MFA, the eigenstates are direct
product of single-site states. In this ground state, the state on
each site is a coherent mixture of the DF state, the vacancy, the
interstitial, and the exciton. This coherence allows constructive
interference between the various hopping processes, thus enhancing
the effect of kinetic energy over that when only vacancies or
interstitials are present or when they hop in a background of only
DF states or only exciton states. This coherent effect is best
understood by examining the highly symmetric case of $\epsilon
_{b}=\epsilon _{a}=\epsilon $, $ t_{b}=t_{a}=t$ and $U=2\epsilon
_{a}$ (so that $\Delta =0^{+}).$ By symmetry, we also have
$\bar{a}=\bar{b}.$ Because of the symmetry between the vacancy and
the interstitial, and between the vacuum and the exciton
states, the MF energy $E_{MF}(\bar{a})$ can be analytically found to be $%
\frac{\epsilon }{2}-\sqrt{\left( \frac{\epsilon }{2}\right) ^{2}+\left( 2zt%
\bar{a}\right) ^{2}}+2zt\bar{a}^{2}$, from which the condition for
the SF state is found to be $\epsilon -2zt<0$. Compared with the
SF condition for vacancy only condensation above the DF
background, we see that the effective hopping integral increases
from $t$ to $2t$ due to the coherent mixture of all 4 states. For
this artificial highly symmetric case, the normal to VI-SF
transition is second order due to $\Delta $ being infinitesimal.
As $\Delta $ increases, the transition becomes first order.  This
is what is shown in Fig 1c as the transition is crossed by
increasing $t.$ Alternatively, the transition can be crossed by
increasing $U,$ as shown in Fig 1(d).

Although our results so far are obtained using the MFA, we believe the
central conclusion, that supersolid can occur even when defects like
vacancies and interstitials have relatively high activation energies is
correct. In the lattice model of (1), the excitations in the supersolid
phase are gapless and the excitations in the normal solid are gapful.
Therefore, the quantum fluctuations are expected to further stabilize the
supersolid phase. In our MFT, there are four supersolid phases in terms of
the order parameters $\bar{a}$ and $\bar{b}$. However, there is no
difference between them in the type of off diagonal ordering of the
underlying $^{4}$He atoms. Thus, they are not really distinct phases in the
sense of different symmetry states, but rather differ in the physics behind
the condensation as discussed above. Indeed, if fluctuations are included or
if we consider non-zero $t_{ab},$ then the sharp distinction between these
''phases'' becomes crossover.

Following Leggett's paper \cite{leggett}, we can derive the SF density
detected by the NCRI experiment, which is
\begin{equation}
\rho _{S}=\rho _{a}m_{0/}m_{a}^{\star }+\rho _{b}m_{0/}m_{b}^{\star },
\end{equation}
where $m_{a}^{\star }$ and $m_{b}^{\star }$ are the effective mass of the
vacancy and interstitial bands respectively determined by the hopping
integral $t_{a}$ and $t_{b}$ and the lattice structure, and $m_{0}$ is the
bare mass of $^{4}$He. For the purpose of illustration, we will assume that $%
m_{b}^{\star }=m_{0}$ at the value of $t_{b}=1/3$. Because the transition at
$T=0$ is first order, the SF density has an abrupt jump at the criticality.
For reasonable values of parameters $\epsilon _{a}=1$, $\epsilon _{b}=4$, $%
t_{a}=0.07$ and $t_{b}=0.2$, we find the jump to be $\sim 9\%,$ which is
about an order larger than the SF density measured in Kim and Chan's
experiments\cite{kim2}. The discrepancy is partly due to the simplicity of
the model. For instance, the critical SF density $\rho _{S}$ is found to be $%
\sim 3\%$ by including an on-site interband hopping $t_{ab}^{0}=0.12$.
Furthermore, the quantum phase fluctuations mentioned above will decrease $%
\rho _{S}$ at criticality by the dual effects of directly
decreasing the order parameters and by reduction of the critical
value $U_{c}$ due to stabilizing of the SF phase relative to the
DF state. We also note that the experiments of Kim and Chan are
performed on granular rather than single crystal, so that the
observed $\rho _{S}$ may be governed by Josephson effect and
therefore can be considerably smaller than the intrinsic value of
homogenous supersolid $^{4}$He.

We have seen that the T=0 transition should be first order. A
careful examination of the finite $T$ transition should include
also the effect of phase fluctuations. The finite $T$ transition
as well as the collective excitations in the various SF phases of
our model will be discussed in a future publication.

In conclusion, we have proposed a solution to how $^{4}$He can become a
supersolid at low $T$ when both vacancies and interstitials have relatively
high activation energies in the normal solid. In our theory, the presence of
low energy bound vacancy-interstitial defects facilitate the condensation of
vacancies. In this theory, the normal state is a commensurate solid while
the supersolid is incommensurate. Furthermore, the local helium density in
the unit cell has different profiles in the two phases. These predictions
can be tested experimentally.

This work is partially supported by the RGC of Hong Kong. We wish to thank
M. H. W. Chan for bringing our
attention to this interesting topic as well as many stimulating discussions.


\begin{thebibliography}{99}
\bibitem{variation}  In these variational calculations shown in Figs.1c, and
1d, for given mean fields $\bar{a}=\bar{b}$, we first solve the mean field
Hamiltonian of (1) to obtain the lowest energy state $\mid \Psi (\bar{a}%
)\rangle $. We then calculate $E(\bar{a})=\langle \Psi (\bar{a})\mid H\mid
\Psi (\bar{a})\rangle $.
\end{thebibliography}
\end{document}